\documentclass[10pt,journal]{IEEEtran}
\usepackage{amsmath}
\usepackage{epsfig}
\usepackage{graphicx}
\usepackage{amsmath}
\usepackage{amsfonts}
\usepackage{latexsym}
\usepackage{amssymb}
\usepackage{cite}
\usepackage{array}

\newcommand{\bz}{\mathbf{z}}
\newcommand{\bd}{\mathbf{d}}

\newcommand{\bw}{\mathbf{w}}

\newcommand{\bi}{\mathbf{i}}
\newcommand{\bp}{\mathbf{p}}
\newcommand{\bY}{\mathbf{Y}}

\newcommand{\bO}{\mathbf{0}}

\newcommand{\bW}{\mathbf{W}}

\newcommand{\bQ}{\mathbf{Q}}

\newcommand{\bg}{\mathbf{g}}

\newcommand{\bI}{\mathbf{I}}

\newcommand{\eps}{\epsilon}

\begin{document}

\title{A Subspace Method for I/Q Imbalance Estimation in Low-IF Receivers}
\author{Ahmad Gomaa$^1$, Ayman Elezabi$^2$ and Mohamed Eissa$^2$ \\
$^1$ Cairo University, Egypt.
$^2$ The American University in Cairo, Egypt}

\maketitle

\begin{abstract}
In low-intermediate-frequency (low-IF) receivers, I/Q imbalance (IQI) causes interference on the desired signal from the blocker signal transmitted over the image frequencies. Conventional approaches for data-aided IQI estimation in zero-IF receivers are not applicable to low-IF receivers. In zero-IF receivers, IQI induces self-interference where the pilots of the image interference are completely identified. However, in low IF receivers, the image interference originates from a foreign signal whose training sequence timing and structure are neither known nor synchronized with the desired signal. We develop a data-aided subspace method for the estimation of IQI parameters in low-IF receivers in the presence of unknown fading channel. Our approach does not require the knowledge of the interference statistics, channel model nor noise statistics. Simulation results demonstrate the superiority of our approach over other blind IQI compensation approaches.
\end{abstract}

\setlength{\textfloatsep}{10pt}
\setlength{\floatsep}{8pt}
\section{Introduction}\label{sec_Introduction}
One of the common impairments of the radio-frequency (RF) frontend is the inphase and quadrature imbalance (IQI) embodied in the gain and phase mismatches between the inphase and quadrature mixers due to manufacturing inaccuracies \cite{Schenk-book}. In the frequency-domain, the impact of IQI appears as interference between the positive and negative frequency components known as the image leakage problem \cite{Schenk-book,Tarighat05}. Consequently, the IQI causes degradation in the signal-to-interference (SIR) power ratio and, hence, the overall receiver performance and throughput \cite{OurTCOM_AF_IQI,ValkamaBlind05,Valkama13-MutInfo}.

Direct (zero-IF) and low-IF down conversion architectures are used in modern receivers to avoid the need for off-chip components \cite{Razavi-book}. Unlike zero-IF, low-IF receivers do not suffer from the DC offset problem. However, in low-IF receivers, neighboring signals (blockers) appear as the image of the desired signal after down conversion. If these blockers leak into the desired signal (due to IQI), they will severely impact the receiver performance especially if they are stronger than the desired signal. Hence, IQI estimation and compensation is vital for low-IF receivers. In \cite{Valkama_LowIF}, the image interference was adaptively and blindly canceled using a single tap IQI compensation method. However, this approach was developed for low input signal-to-interference ratios (SIRs). In \cite{Fettweis_LowIF}, a dual-tap blind IQI compensation filter was proposed based on the time averaging of the observation signals.

In zero-IF receivers, both the desired signal and the IQI-induced interference (image leakage) originate from the transmitted desired signal. Hence, pilots (know signal) are transmitted at both the subcarrier and its image \cite{Minn-Pilot-Design_IQI,Gomaa_IQI_OFDM} to simplify data-aided IQI estimation at the receiver. However, in low-IF receivers, the IQI-induced interference does not originate from the desired signal, and rather comes from a foreign signal. Therefore, the pilots can be received only on the positive frequency components and not on their images since the image frequencies are modulated by a foreign uncontrolled neighboring signal. Accordingly, the conventional data-aided IQI estimation approaches \cite{Gomaa_IQI_OFDM,Pan-IQ-CFO-TWC12,Lin-TCOM10-IQ-CFO} for zero-IF receivers cannot be used for low-IF receivers.

In this paper, we propose a novel subspace-based approach for data-aided compensation in low-IF architectures. We exploit the knowledge of the training sequence at the receiver, construct a data-nulling matrix, and then apply simple algebraic operations to estimate the IQI parameters. The rest of the paper is organized as follows. The system model is introduced in Section \ref{sec_SysModel}, while our data-aided IQI compensation approach is described in Section \ref{sec_algorithm} for flat-fading channels. The extension of our approach for both frequency-selective and time-varying fading channels is also given in Section \ref{sec_algorithm}. The simulation results are given in Section \ref{sec_Simulations}, and the paper is concluded in Section \ref{sec_Conclusion}.

\textit{Notations}: Unless otherwise stated, lower and upper case bold letters denote vectors and matrices, respectively, and $\bI_m$ denotes the identity matrix of size $m$. Also, $\bO$ denote the all-zero vector while $(\,)^*$, $(\,)^T$ and $(\,)^H$ denote the complex conjugate, transpose and conjugate transpose operations, respectively. The notation $\left|\;\right|$ denotes the absolute value.
\section{System Model}\label{sec_SysModel}
\begin{figure}
\centering
\epsfig{file=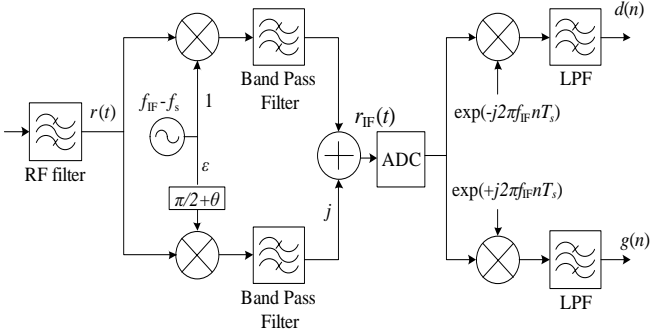,height=1.8in,width=3.4in}
\caption{Low-IF receiver with I/Q imbalance.}\label{fig_sys_model}
\end{figure}

We consider low-IF receivers where the received signal at the output of the RF filter is given by
\begin{align}
r(t)=Re\left\{s(t) \text{exp}\left(j2\pi f_s t\right) + i(t) \text{exp}\left(j2\pi f_i t\right) \right\}
\end{align}
where $Re\left\{.\right\}$ denote the real part. Furthermore, $s(t)$ and $i(t)$ denote the complex baseband equivalent signals of the noisy desired signal and the noisy interference whose carrier frequencies are denoted by $f_s$ and $f_i$, respectively. The source of the interference signal can be a nearby external transmitter or the on-chip transmitter in frequency division duplexing (FDD) systems. Although the RF filter is tuned to $f_s$, its selectivity is not high enough to completely suppress the neighboring interference signals. The reason is that the interference signal is separated from the desired signal by only twice the intermediate frequency $f_\text{IF}$ which is low as the name low-IF suggests. The received signal $r(t)$ is then down-converted into the low IF frequency $f_\text{IF}$ as shown in Fig. \ref{fig_sys_model}. For interference signals located at the image frequency of $f_s$, i.e.,  $f_i = f_s - 2f_\text{IF}$, the IQI-free low-IF signal after the band-pass filters is given by:
\begin{align}
r_\text{IF-IQI-free}(t)= s(t) \text{exp}(+j2\pi f_\text{IF}t) + i(t) \text{exp}(-j2\pi f_\text{IF}t)
\end{align}
where $s(t)$ is easily extracted out through complex digital mixers and digital low-pass filters as shown in Fig. \ref{fig_sys_model}.
However, the analog down-conversion unit suffers from gain and phase mismatches between the paths connecting the oscillator signals to the I and Q mixers \cite{Schenk-book} as shown in Fig. \ref{fig_sys_model}. The gain and phase mismatches are denoted by $\eps$ and $\theta$, respectively. Hence, we write the IQI-impaired complex signal at the analog-to-digital converter (ADC) input as follows \cite{Schenk-book}:
\begin{align}
r_\text{IF}(t)&=\mu\, r_\text{IF-IQI-free}(t) + \nu\, r_\text{IF-IQI-free}^*(t)\nonumber\\
&= \left(\mu\,s(t) + \nu\,i^*(t)\right) \text{exp}(+j2\pi f_\text{IF}t) \nonumber\\
&+ \left(\mu\,i(t) + \nu\,s^*(t)\right) \text{exp}(-j2\pi f_\text{IF}t) \label{eqn_IFsignal}\\
\text{where}\quad &\mu =\frac{1+\eps\,\text{exp}(j\theta)}{2},\qquad \nu =\frac{1-\eps\,\text{exp}(-j\theta)}{2}\label{eqn_IQI_mu_nu}
\end{align}
The conjugate terms ($i^*(t)$ and $s^*(t)$) in \eqref{eqn_IFsignal} appear due to IQI and vanish in IQI-free scenarios where $\eps=1$ and $\theta=0$ yielding $\mu=1$ and $\nu=0$. Next, the low-IF signal $r_\text{IF}(t)$ is digitized and down-converted to DC using digital oscillators and mixers free of IQI as shown in Fig. \ref{fig_sys_model}. The digitally down-converted complex baseband signals are given by
\begin{align}
d(n)&\equiv d(nT_s)= \text{LPF} \left( r_\text{IF}(nT_s) \text{exp}(-j2\pi f_\text{IF}nT_s) \right) \nonumber\\
&= \mu\,s(n) + \nu\,i^*(n) \\
g(n)&\equiv g(nT_s)= \text{LPF} \left( r_\text{IF}(nT_s) \text{exp}(+j2\pi f_\text{IF}nT_s) \right) \nonumber\\
&= \mu\,i(n) + \nu\,s^*(n)
\end{align}
where $T_s$ is the sampling time of the ADC, and $s(n)=s(nT_s)$ and $i(n)=i(nT_s)$ denote the digitized signals of $s(t)$ and $i(t)$, respectively. Furthermore, LPF( ) denotes the complex digital low pass filtering operation with its bandwidth set to that of $s(t)$. In the ideal IQI-free scenario, the signal $g(n)=i(n)$ is not needed because the desired signal solely appears in $d(n)=s(n)$. However, in real-life IQI-impaired systems, the signal $g(n)$ is needed for IQI estimation and compensation as it contains a portion of the desired signal.

\section{Data-aided I/Q Imbalance Estimation}\label{sec_algorithm}
\begin{figure}
\centering
\epsfig{file=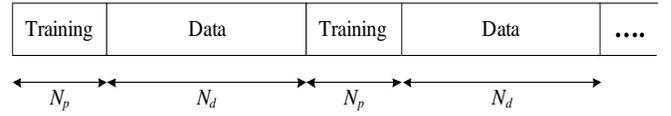,height=0.6in,width=3.4in}
\caption{Frame structure with the training and data sequences.}\label{fig_frame_structure}
\end{figure}

We consider a generic frame structure shown in Fig. \ref{fig_frame_structure} comprising of a length-$N_p$ training sequence followed by a length-$N_d$ data sequence. The training sequence is periodically transmitted for the purpose of channel estimation; however, we exploit it for IQI estimation as well. We collect the samples of $d(n)$ and $g(n)$ over the first training period, $0 \leq n \leq N_p-1$, and construct the following $N_p \times 2$ matrix
\begin{align}
\bY_0\triangleq \begin{bmatrix} \bd_0 & \bg_0 \end{bmatrix} = \begin{bmatrix} \alpha_0\bp+\bz_0 & \bi_0 \end{bmatrix} \begin{bmatrix} \mu & \nu^* \\ \nu & \mu^*\end{bmatrix}
\end{align}
where $\bd_0=\left[d(0) \quad d(1) \quad ... \quad d(N_p-1)\right]^T$, $\bg_0=\left[ g^*(0) \quad g^*(1) \quad ... \quad g^*(N_p-1) \right]^T$, $\bi_0=\left[ i^*(0) \quad i^*(1) \quad ... \quad i^*(N_p-1) \right]^T$. Furthermore, the $N_p\times 1$ complex vectors $\bp$ and $\bz_0$ denote, respectively, the transmitted training sequence and the receiver additive white Gaussian noise (AWGN) at the first training period. We consider frequency-flat fading where the complex fading coefficient over the first training period is denoted by $\alpha_0$ such that $s(n)=\alpha_0 p(n)+z(n)$, $0 \leq n \leq N_p-1$. For the purpose of presentation, we assume that the fading coefficient is static over each training period\footnote{In simulations, the fading coefficients are generated as a time-varying process according to the Jakes' model.}; however, we will show later how to relax this assumption for long training periods. Since we do not have knowledge of the fading coefficient $\alpha_0$, we project the columns of $\bY_0$ on the left-null subspace projection matrix of the known training vector $\bp$ given by
\begin{align}\label{eqn_Qpurb}
\bQ^\perp=\bI_{N_p}-\frac{1}{\bp^H\bp}\bp\bp^H
\end{align}
where $\bQ^\perp\,\bp = \bO$. In order not to change the statistical properties of the noise $\bz_0$, we apply the Gram-Schmidt orthonormalization steps \cite{Golubbook} among the rows of $\bQ^\perp$ as follows:
\begin{align}\label{eqn_Q_orth}
\bQ=orth\left(\bQ^\perp\right)
\end{align}
where $orth(.)$ denotes the well-known Gram-Schmidt operations among the rows of the argument matrix. Then, $\bQ$ is the output $\left(N_p-1\right)\times N_p$ matrix whose rows\footnote{The rank of the projection matrix $\bQ^\perp$ is $N_p-1$, hence it has only $N_p-1$ linearly independent rows.} are normal and orthogonal to each other such that $\bQ\bQ^H=\bI_{N_p-1}$. Note that $\bQ$ is pre-computed and stored at the receiver memory since the training sequence $\bp$ is pre-known. Projecting the columns of $\bY_0$ on $\bQ$, we get
\begin{align}\label{eqn_proj_step}
\bW_0 \hspace{-0.1cm}= \bQ \bY_0 = [\bQ\bz_0 \;\;\, \bQ\bi_0] \begin{bmatrix} \mu & \nu^* \\ \nu & \mu^*\end{bmatrix} \hspace{-0.1cm}\triangleq [ \tilde{\bz}_0 \;\;\, \tilde{\bi}_0]\hspace{-0.1cm} \begin{bmatrix} \mu & \nu^* \\ \nu & \mu^*\end{bmatrix}
\end{align}
where we nulled out the unknown fading coefficient $\alpha_0$. The first and second columns of $\bW_0$ are, respectively, given by
\begin{equation}\label{eqn_w1_w2}
\bw_1 = \mu\,\, \tilde{\bz}_0+ \nu\,\, \tilde{\bi}_0,\qquad
\bw_2 = \nu^* \tilde{\bz}_0+ \mu^* \tilde{\bi}_0
\end{equation}

\noindent From \eqref{eqn_IQI_mu_nu}, we observe that
\begin{align}\label{eqn_mu_nu_relation}
\mu=1-\nu^*
\end{align}
Hence, adding $\bw_1$ and $\bw_2$ yields an IQI-free term as follows:
\begin{align}
\bw_1+\bw_2 = \tilde{\bz}_0 + \tilde{\bi}_0 \label{eqn_added_w1_w2}
\end{align}
Exploiting that $\tilde{\bi}_0$ and $\tilde{\bz}_0$ are statistically uncorrelated, we estimate the IQI parameters as follows:
\begin{align}\label{eqn_mu_times_nu_est}
\widehat{\mu\nu} = \frac{\bw_2^H \bw_1}{\left(\bw_1+\bw_2\right)^H\left(\bw_1+\bw_2\right)} = \nu - |\nu|^2 + e
\end{align}
where $e$ is the estimation error. Nulling out the signal part reduces the power of $\alpha_0\bp+\bz_0$ and, hence, improves the convergence speed of the cross-product term, $\tilde{\bz}_0^H\tilde{\bi}_0$, in the numerator and denominator of $\widehat{\mu\nu}$ in \eqref{eqn_mu_nu_est}. Ignoring the estimation error term $e$ and solving \eqref{eqn_mu_nu_relation} with \eqref{eqn_mu_times_nu_est}, we get
\begin{gather}
\hat{\nu} = \frac{1}{2}-\frac{1}{2}\sqrt{1-4\left(Re\left\{\widehat{\mu\nu}\right\}+\left(Im\left\{\widehat{\mu\nu}\right\}\right)^2\right)} + j Im\left\{\widehat{\mu\nu}\right\} \nonumber\\
\hat{\mu} = 1-\hat{\nu}^* \label{eqn_mu_nu_est}
\end{gather}
where $Im\left\{.\right\}$ denote the imaginary part. An alternative method to estimate the IQI parameters is to rewrite $\bw_1$ as:
\begin{align}
\bw_1 = \frac{\nu}{\mu^*} \mu^*\tilde{\bi}_0 + \mu \tilde{\bz}_0 = \frac{\nu}{\mu^*} \left(\bw_2 - \nu^* \tilde{\bz}_0\right) +\mu \tilde{\bz}_0
\end{align}
Then, we get the least squares estimate of $\nu/\mu^*$ as follows:
\begin{align}\label{eqn_nu_over_mu_est}
\widehat{\left(\frac{\nu}{\mu^*}\right)}=\frac{\bw_2^H\bw_1}{\bw_2^H\bw_2}
\end{align}
Afterwards, we use the relation in \eqref{eqn_mu_nu_relation} and get the estimates of $\mu$ and $\nu$. We found that both methods perform similarly. We use the former method in \eqref{eqn_mu_times_nu_est} and \eqref{eqn_mu_nu_est} in our simulations because it lends itself naturally for future IQI estimation approach combining both data-aided and blind techniques. The averaging in \eqref{eqn_mu_times_nu_est} and \eqref{eqn_mu_nu_est} has been applied in \cite{Fettweis_LowIF} to $d(n)$ and $g(n)$ instead of $\bw_1$ and $\bw_2$ where no projection or linear processing has been applied to $d(n)$ and $g(n)$.
Next, $\hat{\nu}$ and $\hat{\mu}$ are simply used to compensate for the IQI and recover $s(n)$ over the data period as follows:
\begin{align}\label{eqn_IQIcompensation}
\hat{s}(n) = \frac{\hat{\mu}^*\,d(n)-\hat{\nu}\,g^*(n)}{\left|\hat{\mu}\right|^2-\left|\hat{\nu}\right|^2}, \quad N_p \leq n \leq N_d-1
\end{align}
Moreover, we can compensate for IQI over the training period as well and then estimate the channel using one of the standard channel estimation techniques.

\subsection{Time Varying Fading Channels}
If the training period is too long for the fading channel to be considered static, then we divide the training period into $K>1$ segments each of length $L$ such that $KL=N_p$ and $L<N_c$ where $N_c$ is the coherence length (related to the coherence time) over which the fading coefficient is constant. Next, we collect the samples of $d(n)$ and $g^*(n)$ over the $k$-th segment in the vectors $\bd_k$ and $\bg_k$, respectively. Then, we apply the projection step to each of the $K$ segments to get the matrices $\left\{\bW_0^k=[\bQ_k\bd_k\;\bQ_k\bg_k],\,0\leq k \leq K-1\right\}$ each of size $(L-1) \times 2$, where the $k$-th projection matrix $\bQ_k$ is computed as in \eqref{eqn_Qpurb} and \eqref{eqn_Q_orth} but for the $L$-length $k$-th pilot segment. Vertically concatenating these $K$ matrices, we get the following $(N_p-K) \times 2$ matrix
\begin{align}
\bW\triangleq\begin{bmatrix}\bW_0\\ \vdots \\ \bW_{K-1} \end{bmatrix} &= \begin{bmatrix}\bQ_0\bz_0 & \bQ_0\bi_0 \\ \vdots & \vdots \\ \bQ_{K-1}\bz_{K-1} & \bQ_{K-1}\bi_{K-1} \end{bmatrix} \begin{bmatrix} \mu & \nu^* \\ \nu & \mu^*\end{bmatrix} \hspace{-0.1cm} \nonumber\\
&\triangleq [\bw_1\;\bw_2] \begin{bmatrix} \mu & \nu^* \\ \nu & \mu^*\end{bmatrix}
\end{align}
Note that $\bW$ has the same structure as $\bW_0$ in \eqref{eqn_proj_step} since $\mu$ and $\nu$ are assumed constant over the $K$ segments. Next, we estimate the $\mu$ and $\nu$ as in \eqref{eqn_mu_times_nu_est} and \eqref{eqn_mu_nu_est}. Since $K>1$, the lengths of $\bw_1$ and $\bw_2$ decrease by $K-1$ from the scenario in Section \ref{sec_algorithm} where the channel is static over the whole training period of length $N_p$. Hence, the averaging gain (embodied in the dot products) in \eqref{eqn_mu_nu_est} decreases.

In a similar approach, we separately apply the projection method to the next training periods and vertically concatenate the resulting matrices to increase the dimension of $\bW$ and, hence, the averaging gain. This helps reducing the LSE error and increasing the accuracy of the IQI estimates.

\subsection{Note About Frequency Selective Fading Channels}
Although we explained our approach for frequency-flat fading channels, we can easily extend it to frequency selective channels by working in the frequency-domain (FD) instead of the time-domain. To avoid redundant equations, we describe the extension in words as follows. It is well known that frequency selective channels are converted into a set of flat fading channels \cite{OFDM-book} by applying the discrete Fourier transform to $d(n)$ and $g(n)$. Instead of exploiting the time coherence for flat fading channels, we exploit the frequency coherence for frequency selective channels. We group the frequency neighboring pilots into one segment over which the channel frequency response is quasi-static. Then, we apply the projection method over each segment and proceed as in \eqref{eqn_proj_step}-\eqref{eqn_mu_nu_est}. IQI is easily compensated in the FD by linearly processing the subcarrier and its image similar to \eqref{eqn_IQIcompensation}. If the data and pilot symbols are frequency-multiplexed \cite{LTE}, then the pilots should be lumped next to each other in the FD for our approach to work. This is different from the data-pilot structure for IQI compensation in zero-IF receivers where the pilots are placed in the image frequencies of each other \cite{minn_pilot}.
\begin{figure}
\centering
\epsfig{file=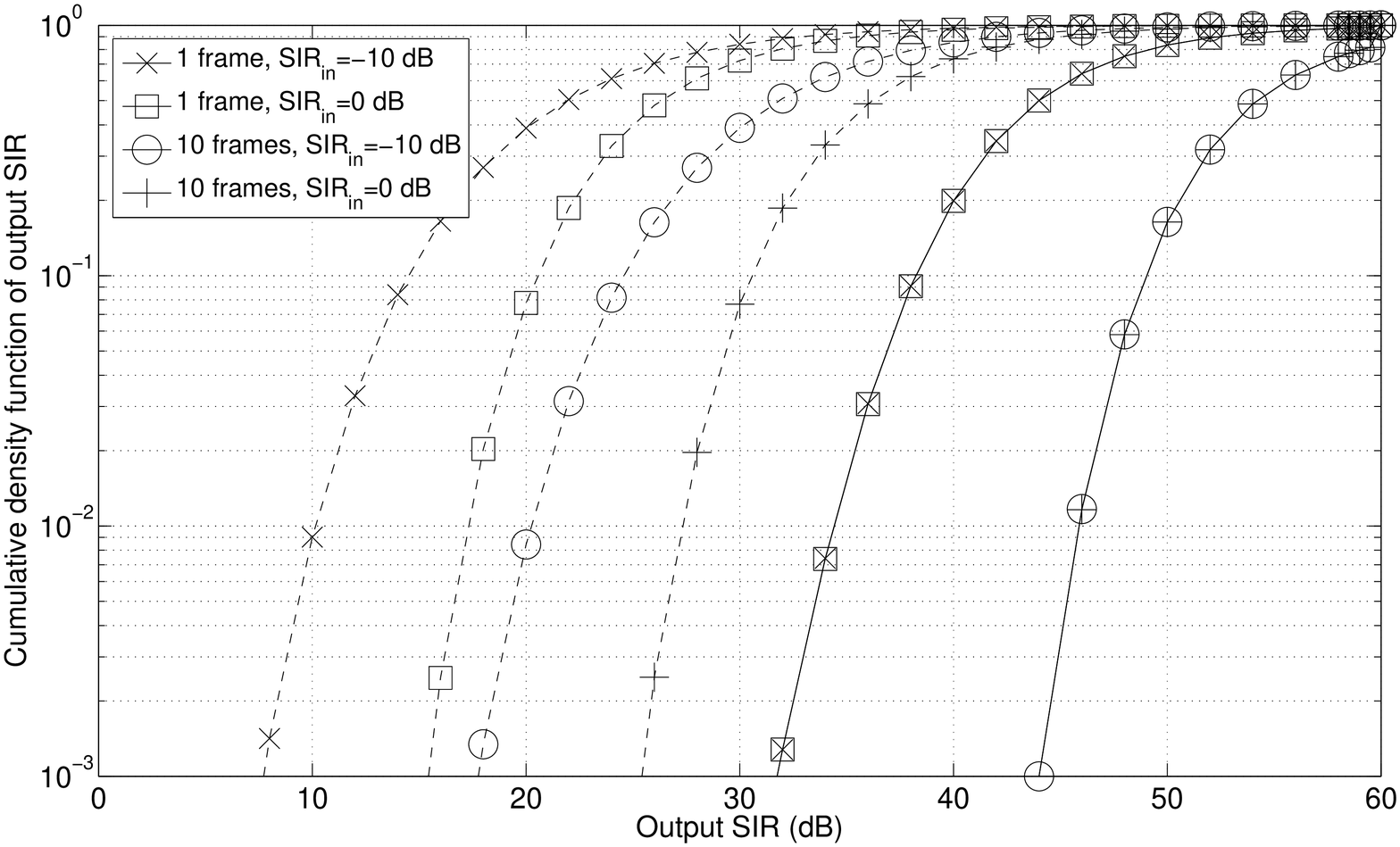,height=2.4in,width=3.4in}
\caption{Output SIR CDFs of data-aided (solid lines) and blind (dashed lines) approaches at SNR = 35 dB for different $\text{SIR}_\text{in}$ levels and number of frames.} \label{fig_CDF_diff_SIR_frames}
\end{figure}

\section{Simulation Results}\label{sec_Simulations}
We simulate the performance of our proposed IQI estimation approach for the frame structure in Fig. \ref{fig_frame_structure} where the training and data sequences are time-multiplexed with $N_p = 8$ and $N_d = 48$. The data period is chosen to be six times that of the the training similar to the 3GPP long term evolution (LTE) uplink standard \cite{LTE} where any consecutive training symbols are separated by six (five) data symbols for normal (extended) cyclic prefix. In the simulations, we use the time-varying Rayleigh flat fading channel model generated according to Jakes' model with the doppler frequency $F_D = 100$ Hz and sampling time $T_s = 2 $ micro seconds. These parameters correspond to a system bandwidth of 250 kHz, carrier frequency of 1.5 GHz, and realtive transmitter-receiver speed of 45 miles per hour. We use gain and phase imbalances of $20\,\text{Log}_{10} \eps$ = 1 dB and $\theta=2^\circ$. The signal-to-noise ratio (SNR) is given by SNR=$E_s/N_o$ where $E_s$ is the average received signal power and $N_o$ is the single-sided power spectral density of the receiver AWGN. The training sequences are constructed using the Zadoff-Chu sequence \cite{Chu72} while the data sequences are transmitted using the 64-QAM single-carrier modulation technique. We compare our algorithm with the blind approach in \cite{Fettweis_LowIF}. In Fig. \ref{fig_CDF_diff_SIR_frames}, we plot the cumulative density function (CDF) of the output SIR after 1 and 10 frames for different input SIR levels defined by $\text{SIR}_\text{in}=E\left[|s(n)|^2\right]/E\left[|i(n)|^2\right]$. The CDF is the probability that the output SIR is smaller than $x$ with $x$ given on the horizontal axis. As expected, increasing the number of frames improves the performance of both data-aided and blind techniques thanks to the increased number of averaging symbols.
\begin{figure}
\centering
\epsfig{file=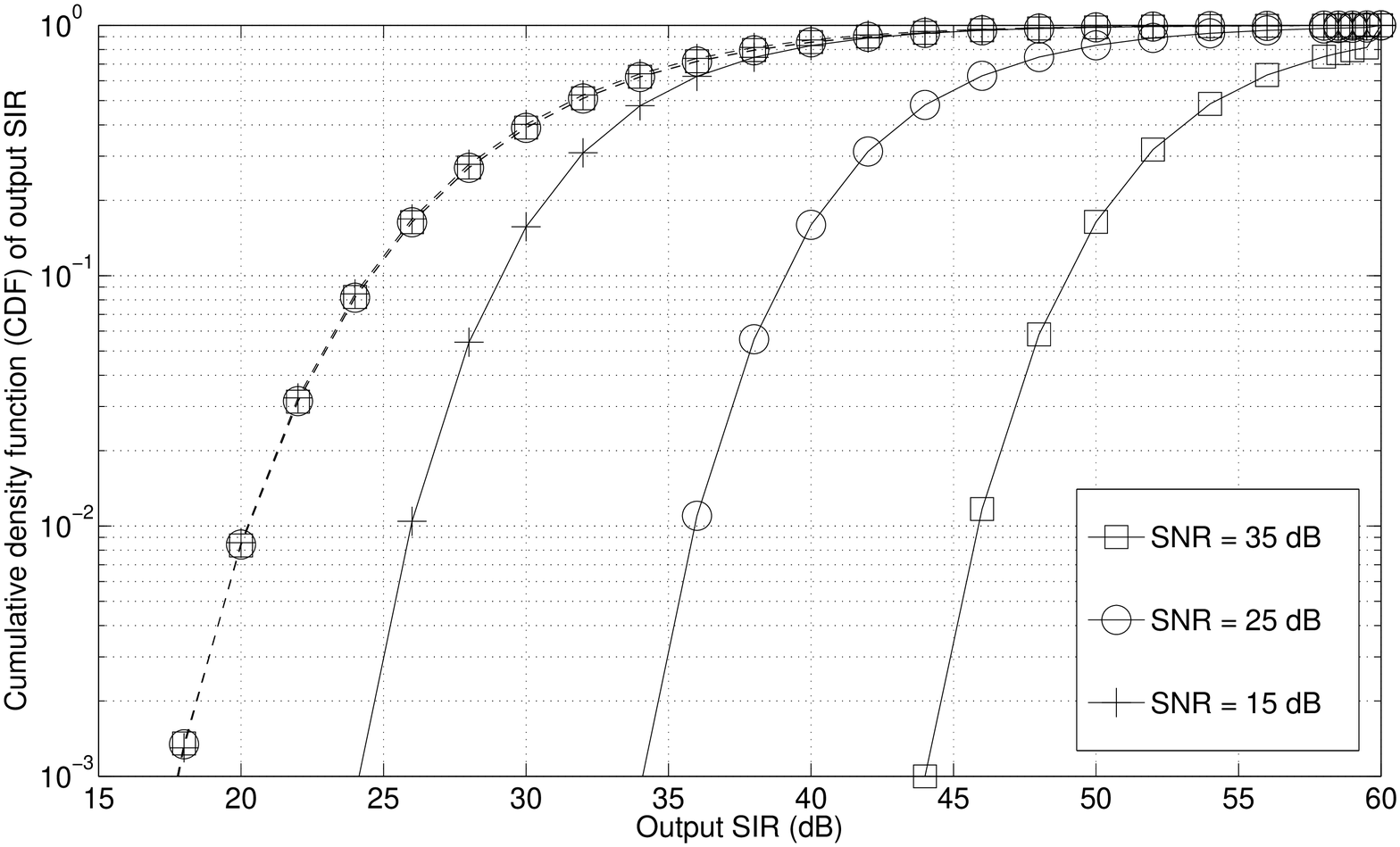,height=2.4in,width=3.4in}
\caption{Output SIR CDFs of data-aided (solid lines) and blind (dashed lines) approaches for different SNR levels after 10 frames at $\text{SIR}_\text{in}$ = - 10 dB} \label{fig_CDF_diff_SNR}
\end{figure}
However, increasing the interference power by 10 dBs severely impacts the blind technique while it does not affect our data-aided technique. In our approach, the signal is nulled out leaving us with small amount of noise in high SNR scenarios. In the extreme case of noise-free systems, we block the signal leaving only the interference, i.e., $\tilde{\bz}_0=0$ in \eqref{eqn_w1_w2}. Hence, only one time sample is sufficient for perfect IQI estimation regardless of the interference power, c.f. \eqref{eqn_mu_times_nu_est} and \eqref{eqn_nu_over_mu_est}. In Fig. \ref{fig_CDF_diff_SNR}, we compare the data-aided and blind approaches for different SNR levels. Unlike the blind approach, our data-aided technique is improved by increasing the SNR because it exploits the pilot knowledge to knock the strong signal out.

\section{Conclusion}\label{sec_Conclusion}
We proposed a novel data-aided subspace-based approach for IQI parameters estimation in low-IF receivers in the presence of unknown fading channel. Our solution is based on projecting the received signal on the left null subspace of the known training sequence to get rid of the unknown signal and channel. We showed how to apply our approach to both frequency-flat and frequency-selective channels. We considered the scenario where data and pilot symbols are time-multiplexed, and further developed the conditions on the pilots locations for frequency-multiplexed data and pilots. Simulation results showed the effectiveness of our approach as compared to blind IQI compensation approaches at different number of frames and several SNR and SIR levels. Our future work will focus on the design of combined data-aided and blind IQI estimation techniques.

\bibliographystyle{IEEEtran}
\bibliography{IEEEabrv,reference}
\end{document}